\documentclass[10pt]{IEEEtran}

\usepackage{amssymb,amsthm,mathrsfs,mathtools}
\usepackage[normalem]{ulem}

\usepackage{framed}
\usepackage{wrapfig}
\usepackage{algorithmic}
\usepackage{algorithm}
\algsetup{linenosize=\scriptsize}
\usepackage{amsmath}
\usepackage[dvipsnames]{xcolor}
\usepackage{cite}
\usepackage{url}
\usepackage[final]{graphicx}
\usepackage{subfigure}
\graphicspath{{./}{./figures/}}
\renewcommand{\figurename}{Fig.}

\newcommand{\comment}[1]{ }

\renewcommand{\algorithmicrequire}{\textbf{Input:}}

\newcounter{tc}

\newcommand{\myitemizebegin}{\begin{list}{$\bullet$}
{
 \setlength{\leftmargin}{0.4cm}
 \setlength{\parsep}{0.0cm}
 \setlength{\itemsep}{0.05cm}
 \setlength{\topsep}{0.0cm}
}}
\newcommand{\myitemizeend}{\end{list}}

\newcommand{\myitemizebeginless}{\begin{list}{$\bullet$}
{
 \setlength{\leftmargin}{0.7cm}
 \setlength{\parsep}{0.0cm}
 \setlength{\itemsep}{0.05cm}
 \setlength{\topsep}{0.0cm}
}}
\newcommand{\myitemizeendless}{\end{list}}

\newcommand{\SU}{\ensuremath {\mathit{SU}}}

\newcommand{\proposed}{{\tt {Weighted-CSR}}}

\newcommand{\q}{{\tt {q}}}

\newcommand{\J}{{\tt {u}}}
\newcommand{\I}{{\tt {n}}}

\title{Dynamic Spectrum Sharing in the Age of Millimeter Wave Spectrum Access}

\author{Bechir Hamdaoui$^*$, Bassem Khalfi$^{\dag}$, Nizar Zorba$^{\ddag}$
\thanks{An IEEE-formatted version of this article is published in IEEE Network. Personal use of this material is permitted. Permission from IEEE must be obtained for all other uses, in any current or future media, including reprinting/republishing this material for advertising or promotional purposes, creating new collective works, for resale or redistribution to servers or lists, or reuse of any copyrighted component of this work in other works.}
~\\
$^*$ \small Oregon State University, \small Corvallis, hamdaoui@oregonstate.edu~\\
$^{\dag}$ \small Qualcomm, \small San Diego, bkhalfi@qti.qualcomm.com~ \\
$^{\ddag}$ \small Qatar University, \small Doha, nizarz@qu.edu.qa~
}

\IEEEoverridecommandlockouts
\twocolumn

\begin{document}

\maketitle

\begin{abstract}
Next-generation wireless networks are facing spectrum shortage challenges, mainly due to, among other factors, the projected massive numbers of IoT connections and the emerging bandwidth-hungry applications that such networks ought to serve. Spectrum is scarce and expensive, and therefore, it is of crucial importance to devise dynamic and flexible spectrum access policies and techniques that yield optimal usage of such a precious resource. A new trend recently being adopted as a key solution to this spectrum scarcity challenge is to exploit higher frequency bands, namely mmWave bands, that were considered impractical few years ago, but are now becoming feasible due to recent advances in electronics.
Though, fortunately, spectrum regulatory bodies have responded by allowing the use of new bands in the mmWave frequencies, much work still needs to be done to benefit from such new spectra.

In this paper, we discuss some key spectrum management challenges that pertain to dynamic spectrum access at the mmWave frequencies, which need to be overcome in order to promote dynamic spectrum sharing at these mmWave bands. We also propose new techniques that enable efficient dynamic spectrum sharing at the mmWave bands by addressing some of the discussed challenges, and highlight open research challenges that still need to be addressed to fully unleash the potential of dynamic spectrum sharing at mmWave bands.
\end{abstract}

\section{Introduction}
Emerging 5G wireless networks are anticipated to bring about a technological transformation in modern societies by providing ultra-reliable, high-speed communication infrastructures, projected to  serve billions of wireless devices, machines and vehicles.
%
Evolving next-generation wireless technology targets several performance requirements, including low latency, high mobility, high connection density, and high throughput~\cite{saadat2018multipath}.
Such stringent requirements give rise to several challenges that next-generation networks need to overcome, and of particular interest to this paper is that of spectrum/bandwidth shortage, which desperately calls for innovative dynamic spectrum access and management approaches, that can make efficient use of the available spectrum to meet the projected performance targets.

The already congested sub-6 GHz band does not have any extra bandwidth to offer, thereby prompting spectrum regulatory bodies and technology providers to shift toward millimeter Wave (mmWave) band access. These mmWave bands are attractive because they (among other factors): i) provide higher amounts of bandwidth, ii) have small wavelengths, allowing the use of higher numbers of transmit/receive antennas, essential to enable the massive Multiple Input Multiple Output (mMIMO) technology.

This paper discusses key challenges that the dynamic spectrum sharing paradigm faces when considering mmWave band access, and propose new solution approaches that aim to overcome these challenges. In this paper, we also highlight open research challenges that remain to be addressed, for the integration of dynamic spectrum sharing at mmWave bands in next-generation standards. We will specifically tackle the 5G standard as the current state-of-the-art wireless network, but the challenges and proposals within this paper would apply to any next-generation network as well.


\subsection{5G's Key Dynamic Spectrum Sharing Challenges}
\label{subsubsec:challenges}
To meet the high bandwidth requirement of next-generation wireless systems, spectrum regulators worldwide have opened up new bands for usage. For example, FCC has recently established new rules for allowing mmWave band use for wireless broadband devices in frequencies above 24 GHz. Likewise, Ofcom has adopted similar rules for frequencies above 30 GHz.
Recognizing the importance of exploiting the newly available wideband spectrum, 5GPP, for instance, has been focusing on promoting flexible radio access, and adopting dynamic spectrum sharing for 5G systems to enable efficient access to wideband spectrum, thus increasing spectrum usage efficiency and addressing bandwidth shortages~\cite{li2016qos}.
To ease the promotion and adoption of such technologies, there remains, however, some key challenges that one needs to address, thus achieving successful dynamic spectrum sharing in these mmWave bands.


\myitemizebegin

\item {\bf Spectrum usage heterogeneity challenge:} Multiple different types of 5G devices, running various applications with different QoS requirements and traffic demands, will be connected to and supported by the 5G network infrastructure. This diversity will result in heterogeneous access to the 5G's wideband spectrum, where different bands will experience different occupancy and usage behaviors with variability across time, frequency, and space.

\item {\bf Ultra-wideband (sub-6GHz \& mmWave) spectrum access challenge:} The spectrum bands opened up for 5G span a wide range of frequencies, ranging from low-frequency ($\leq$ 1 GHz) to high-frequency (mmWave/above-24 GHz) bands. The wideband nature renders existing narrowband spectrum sensing approaches unsuitable for mmWave
    sensing.
    For instance, signals transmitted at mmWave frequencies are subject to severe attenuation, leading to high false positives of spectrum occupancy information. Therefore, calling for cooperative sensing approaches that can overcome the hidden terminal problem, where devices located further away from the signal source may not be able to sense primary signals. In addition, mmWave spectrum is in the order of GHz whereas low-frequency narrowbands are in the order of MHz,
    thereby calling for new sensing approaches that are suitable for mmWave spectrum access.

\item {\bf Limited hardware capability challenge:}
   Wideband spectrum usage varies over time and from one frequency band to another, making existing spectrum availability recovery techniques---which, in most part, leverage spectrum occupancy variability to provide efficient recovery---suffer from current hardware limitations. This leads to inaccurate and unreliable recovery of spectrum availability information.

\myitemizeend

\subsection{Objective of this Work}
We address the aforementioned challenges by proposing mmWave spectrum availability recovery techniques that:
\myitemizebegin
\item {\em {Overcome the high sampling rate requirement}} that naturally arises when sensing signals at mmWave frequencies in a large bandwidth. Traditional spectrum availability recovery approaches are not suitable for the high and wide frequency range of mmWave bands, as they would require costly hardware, and sophisticated DSP algorithms to be able to recover occupancy at such high frequencies. We present new techniques that will take advantage of the spectrum usage heterogeneity naturally inherent to wideband access, as well as user cooperation to make substantial reduction in mmWave spectrum recovery overheads, without compromising sensing accuracy.
\item {\em {Overcome tempo-spectral sparsity variability}} that is inherent to wideband spectrum occupancy behavior. Spectrum availability sparsity depends on spectrum occupancy, and hence, the time variability of the occupancy makes the sparsity level also time varying. We introduce techniques that adapt in realtime to sparsity variability, without compromising recovery accuracy and sensing overhead.
\item {\em {Mitigate the increased signal propagation loss at mmWave frequencies}} while maintaining low sensing overheads.
Our techniques achieve this objective by leveraging collaborative filtering theory, and taking advantage of the high user density requirement of 5G to reduce false positives of spectrum availability information.
\myitemizeend

\section{Compressed MmWave Spectrum Availability Recovery}
\label{sec:priorwork}
Spectrum sensing is essential to the promotion of dynamic wideband spectrum sharing. Unlike narrow-band sensing, the challenge with wideband is that it requires high sampling rates, which can incur significant energy, computation, and communication sensing overheads.
Motivated by the sparsity nature of spectrum occupancy, researchers have put significant efforts on exploiting compressed sampling theory to make mmWave spectrum sensing possible at sub-Nyquist rates (see~\cite{sun2013wideband} for a survey on the subject).
A common factor in most prior works is the assumption that the sparsity level is fixed over time~\cite{hamdaoui2018compressed}.
Several works exploit additional signal information knowledge to improve spectrum availability recovery by, for instance, formulating the sensing task as an $\ell_1-$minimization optimization, that exploits knowledge about the support of the sparse signal, to recover information from noisy measurements~\cite{needell2016weighted}.
The signal support here refers to the set of sub-bands of the wideband spectrum that are occupied by some primary users, and hence, when spectrum occupancy is sparse, the support of the wideband is small.
These approaches, however, work well in applications where the support of the spectrum does not change much over time~\cite{needell2016weighted}.
In wireless communication applications using mmWave/wideband access, the signal support, however, does change over time, and estimating the support is too difficult to do in advance, thereby limiting the suitability of these approaches within mmWave/wideband access.
%

\subsection{Weighed Compressed Spectrum Availability Recovery}
One observation we make by investigating existing compressive spectrum sensing approaches is that they consider wideband spectrum occupancy to be {\em homogenous}, meaning that the mmWave spectrum is considered as one single multi-band block, whose sparsity level---number of occupied bands---is considered to be the same for the entire spectrum. However, because applications of similar types (TV, satellite, cellular, etc.) are often assigned bands within the same block, different blocks exhibit different occupancy patterns. Hence, wideband spectrum occupancy may vary signiﬁcantly from one block to another, a trend that has also been confirmed by real measurement studies~\cite{yilmaz2016determination}.
Incorporating this fine-grained sparsity structure into the formulation of wideband spectrum occupancy recovery is recommended, as it improves the recovery performance and enhances the detection accuracy.
%
The novelty of the proposed idea lies then in exploiting the fact that wideband spectrum tempo-spectral occupancy is {\em time-varying} and {\em heterogeneous}, where different frequency blocks exhibit different band occupancy behaviors, with band occupancies varying considerably over time, and from one frequency block to another.
%
%
Our proposed approach, presented next, exploits this {\em block-like tempo-spectral occupancy structure}, to improve occupancy recovery of wideband access.

Without loss of generality, let us assume that the available wideband spectrum is divided into $n$ narrowbands, and consider a primary signal that is occupying the entire wideband.
Our goal is to propose an efficient sensing approach that allows a secondary user (\SU), receiving the primary signal, to recover the occupancy information of the $n$ narrowbands.
Recall that compressive sampling (CS) theory exploits the spectrum occupancy sparsity to enable recovery of the spectrum occupancy vectors from only smaller numbers of measurements.
Various CS-based approaches have been proposed in the literature ranging from optimization-based (e.g., LASSO) to greedy-like (e.g., OMP and CoSaMP) approaches~\cite{davenport2011introduction}.
%

Let's first assume that the $n$ narrow-bands are grouped into $g$ disjoint contiguous blocks,
with each block consisting of $n_i$ contiguous bands.
The block-like structure of spectrum occupancy behavior dictates that the average number of bands occupied within a block varies from one block to another; when necessary, blocks with similar sparsity levels are merged together and assigned a sparsity level that corresponds to their overall average.
Note that, in practice, different blocks could exhibit quite different occupancy averages, where a block corresponds to a set of bands assigned to the same application type~\cite{yilmaz2016determination}. These averages are often available via measurements studies~\cite{yilmaz2016determination}, or could also possibly be provided by spectrum operators if appropriate incentives are given.
Our recovery approach, referred to as {\proposed} throughout, consists of exploiting the sparsity variability observed across the different frequency blocks to allow for a more efficient solution search.
To illustrate this further, we formulate a weighted $\ell_1-$minimization where the weights assigned to eacg block of the spectrum are defined as inverse proportional to the average occupancy. This way, the weights are designed in such a way that a block with higher sparsity is assigned a smaller weight.
The detailed formulation and more insights are further discussed in~\cite{hamdaoui2018compressed}, where the key result is that assigning smaller weights to blocks with higher sparsity levels forces the search for a sparse solution vector to be geared to lesser sparse blocks, which improves sensing in terms of both recovery error and measurement overhead~\cite{khalfi2018efficient}.

\subsection{Error Performance Analysis}
The solution obtained under \proposed~ yields lower error than the solution obtained from LASSO~\cite{candes2006stable}, 
with an overwhelming probability (the probability expression is provided in~\cite{khalfi2018efficient}).
%
\begin{figure}
	\center{
	\includegraphics[width=1\columnwidth]{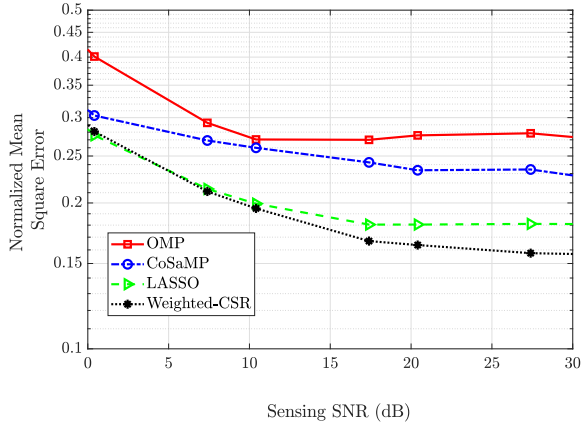}
	\caption{\small{Error performance.}}
	\label{fig:perfSNR}}
\end{figure}
In Fig.~\ref{fig:perfSNR}, we also show that \proposed~outperforms not only LASSO, but also OMP and CoSaMP~\cite{davenport2011introduction}.
%
Compared to LASSO, \proposed~yields higher accuracy, without incurring extra computation~\cite{khalfi2018efficient}.
However, this accuracy gain comes at the cost of needing to know occupancy information in advance (the average occupancy of each block). When compared to the heuristic OMP and CoSaMP approaches, \proposed~has significantly higher accuracy, but requires more computation~\cite{khalfi2018efficient}.

\subsection{Sensing Overhead Performance Analysis}
The sensing overhead is mainly depending on how many samples are needed to be able to detect the occupied bands, in the wideband spectrum of interest. Given the heterogeneous spectrum occupancy, the sensing matrix can be designed in a non-uniform manner, so that the needed number of measurements $m$ is lower than the conventional way. A bound on the required number of measurement is derived in~\cite{khalfi2018efficient}, which is shown to be function of the number of bands, the average occupancy of the wideband, and the average occupancy of each block of the wideband spectrum. Having a lower number of measurements reduces the sensing overhead.


%
\begin{figure}
	\center{
	\includegraphics[width=1\columnwidth]{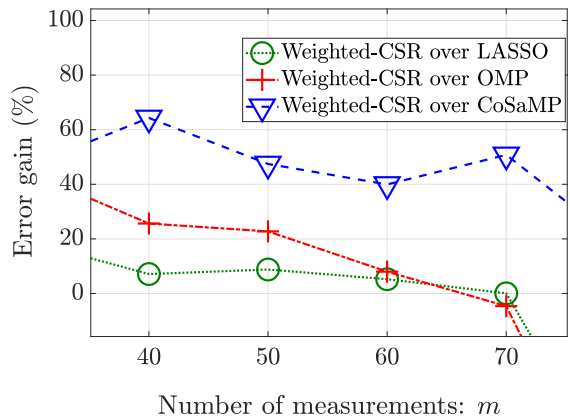}
	\caption{\small{Sensing overhead.}}
    \label{fig:perfm}}
\end{figure}
%
%
Fig.~\ref{fig:perfm} shows that for a given number of measurement $m$, \proposed~incurs an error that is smaller than that incurred by any of the other three studied approaches. This could also be interpreted as follows: for a fixed error target, the proposed \proposed~requires lesser numbers of measurements (and hence incurs lesser sensing overhead) when compared to LASSO, OMP and CoSaMP.

\section{Open MmWave Spectrum Sharing Challenges}
In the previous section, we showed that the proposed \proposed~approach outperforms its state-of-the art counterpart techniques by improving spectrum occupancy recovery accuracy, and reducing sensing overhead considerably. However, there remains key practical challenges that need to be addressed to unleash the full potential of \proposed~for mmWave spectrum access.

\subsubsection {Time-varying occupancy} Spectrum occupancy sparsity is time-varying, and hence so is the number of required measurements, $m$. Therefore, recovery approaches should adopt $m$ to the instantaneous sparsity in real time, so as occupancy information can be recovered with minimum overhead.

\subsubsection {Limited hardware} The number of branches that state-of-the art hardware architectures can support is way smaller than the required value of $m$. This calls for new techniques that overcome this hardware limitation, so that the recovery process still maintains its high accuracy performance, but without incurring extra sensing delay and/or energy overheads.

\subsubsection {Fading and limited sensing sensitivity} Wireless channel fading and limited sensitivity of radio transceivers may lead to inaccurate detection of the primary signals, resulting in increased false-positive rates of the spectrum occupancy information.

\subsubsection {Increased signal path loss} The increased signal path loss, that naturally arises from the transmissions at mmWave frequencies, can lead to high false positives of spectrum occupancy information. This is because transmitting at high frequencies intensifies the hidden terminal problems, where devices may not be able to detect/sense the presence of primary signals sent at high frequencies, leading to unreliable recovery of spectrum occupancy.

\section{Towards Practical Spectrum Availability Recovery for MmWave Band Access}
We now present two practical solutions that aim to address the above challenges.
\subsection{Cooperative \proposed~for MmWave Access}
In this first approach, we leverage user cooperation to enable \proposed~while overcoming three of the aforementioned practical challenges: the time-varying nature of spectrum occupancy, transceiver hardware limitation, and fading and sensing sensitivity challenge.

\subsubsection{Proposed Solution}
As argued earlier, the number of measurements, $m$, needed to perform the \proposed~recovery is time varying, since the sparsity of the tempo-spectral spectrum occupancy changes over time.
Furthermore, the maximum number of streams a radio transceiver can have is way smaller than typical values of $m$; for instance, for a 1 GHz wideband spectrum with 20 MHz band resolution, a value of at least $m=16$~\cite{yazicigil2015wideband} is needed, whereas the number of streams can only be of the order of 3 or 4. This hardware limitation makes it impossible to perform \proposed~in only one hardware scan. In an effort to overcome this hardware limitation, recent works have resorted to sequential sensing/scanning, which essentially generates the $m$ measurements via multiple, sequential scans using the same hardware~\cite{Yazicigil2017how}. Though overcomes the hardware issue, this approach incurs excessive delays, rendering the approach unsuitable for realtime applications.
Hybrid precoding architectures are other techniques for mmWave access that are shown to reduce the number of required RF chains compared to the number of antennas, so as to reduce the total number of needed data streams~\cite{chen2018non}. The idea essentially consists of exploiting the mmWave spectrum access sparsity inherently present in the angular domain, to construct novel non-uniform quantization (NUQ) codebooks, that map the quantization bits non-uniformly for different coverage angles. Although this novel technique reduces the number of RF chains compared to state-of-the art approaches, the value of $m$ required by \proposed~is still higher than what these novel architectures allow.

In this work, we leverage cooperation among \SU s to overcome this issue. But before delving into the proposed idea, it is worth noting that cooperative approaches have already been studied extensively in the literature to address various issues (e.g. to address the classical {\em hidden terminal problem}), and thus, we do not claim to propose cooperative techniques in the work. In effect, we assume that cooperation is already adopted in these sensing schemes for the purpose of addressing hidden terminal problem, arising from the spatial distribution of users, and we will leverage such cooperative mechanisms to address the limitation in the number of RX streams, as well as the time-variability of signals' sparsity.
Specifically, instead of having each \SU~sequentially take and report $m$ sensed measurements~\cite{Yazicigil2017how}, our proposed technique consists of having all \SU s each (in parallel) perform one sensing scan using its available (limited) number of branches and send its partial measurement vector to a fusion center---a fusion center's role could be played by any \SU~or any other independent entity.
When the number of combined measurements received by the fusion center reaches $m$, the fusion center applies \proposed~to recover the spectrum availability.
Note that an \SU~can choose to perform multiple (sequential) sensing scans via its hardware to make the $m$ needed measurements, but this will lead to higher sensing delays.

\begin{figure}
	\center{
		\includegraphics[width=1\columnwidth]{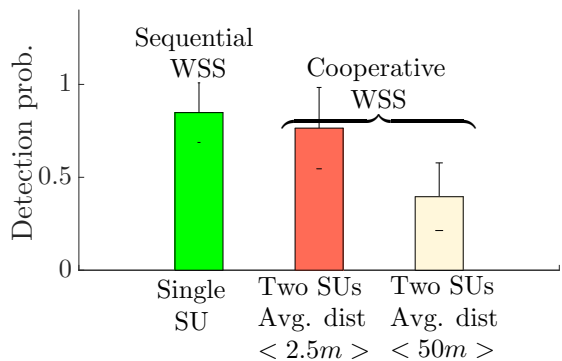}
		\vspace{-0.35in}
		\caption{\small{Benefits of cooperation. 'Sequential Wideband Spectrum Sensing (WSS)' refers to when spectrum sensing is done sequentially, whereas 'Cooperative WSS' refers to our proposed recovery approach.}}
		\label{fig:comp}}
\end{figure}
The challenge with this presented approach, however, is that because of channel fading, each of the \SU s participating in the cooperative sensing task observes a different occupancy vector, resulting in inconsistent measurements across the \SU s. To address this challenge, we propose to consider only close-by users to participate in the cooperative sensing task for that specific zone, so that their observed vectors are likely to be similar.

%
%
Clearly, the higher the gap between the observed vectors of close-by users, the higher the noise variance, and hence, the worse the recovery accuracy.
Fig.~\ref{fig:comp} presents simulation results that support this observation. Our simulations consider two close-by \SU s, a wideband system with $n=200$ and $m=16$, and a fading environment. 
Note that cooperation achieves a detection probability that is as high as that obtained under the recently proposed sequential sensing approach~\cite{Yazicigil2017how}. Also, observe that as the difference between the received signals (i.e., distance between \SU s) becomes high, the accuracy of the recovery drops drastically, due to inconsistency in the observed signals.

In addition to overcoming the hardware limitation as explained above, cooperative \proposed~can address the time-variability challenge of the sparsity level due to time-varying occupancy. Since $m$ depends on the spectrum occupancy, and this occupancy is time-varying, then so is $m$. Fortunately, cooperation can also help overcome the variability of $m$ in time. Let's for example assume that $m$ varies from $10$ to $15$, and each \SU~can only collect 5 measurements (due to hardware limitation). Without cooperation, one \SU~cannot recover spectrum availability, since $5<m$. However, with $3$ cooperative \SU~each collecting $5$ measurements, we can have a total of $15$ measurements, which is required to recover the spectrum occupancy. If, at a later time, $m$ increases to $20$, then we can consider an extra cooperative \SU~to collect $5$ extra measurements, making up the $20$ needed measurements. Therefore, cooperation can also be leveraged to address the time-variation of the number of measurements, $m$.

\subsubsection{Further Research Challenges}
One key observation that we make here is that although fading/distance makes each \SU~observe a different signal, all observed signals are likely to have the same vector support. Therefore, one can take advantage of this observation and develop distributed approaches that combine measurements received from the different users to recover the common vector support.
Our approach is to first consider techniques that rely on measurements with low variation; e.g., by only allowing close-by \SU s to participate in the cooperation. Another issue that is worth further investigation is to study the delay/accuracy tradeoffs between the proposed cooperative approach and that of the sequential sensing approach~\cite{Yazicigil2017how}. While sequential sensing incurs delays due to its need for using the hardware multiple times to generate the $m$ required measurements, cooperation also incurs overhead in the process of sharing sensing reports among users. However, our argument is that cooperation will already be needed for addressing hidden terminal issues, and hence, we are only leveraging it too, for the sake of addressing our sensing issues.
%

\subsection{Robust mmWave Spectrum Availability Recovery Under Increased Path Loss}

In Section~\ref{sec:priorwork}, we showed that by exploiting the {\em sparsity} and {\em heterogeneity} of spectrum occupancies, an \SU~can recover wideband occupancy information at sub-Nyquist rates, yielding substantial reduction of sensing overheads. In the previous section, we discussed how leveraging cooperative sensing can overcome the hardware limitation, as well as the time-varying nature of spectrum occupancy/sparsity. In this section, we present a technique that addresses issues arising from the increased propagation loss at mmWave frequencies, and we do so by leveraging low-rank matrix approximation theory (aka collaborative filtering in the machine learning community)~\cite{candes2009exact} to devise approaches that {reduce false positives of spectrum occupancies at mmWave bands,} in the presence of high signal strength decay, without incurring higher sensing overheads.


The key observation that enabled the use of low-rank matrix theory is that in wideband spectrum access, the occupancies of the spectrum bands observed by \SU s located within the same region are very likely to be (roughly) the same for some set of bands (depending on how close the \SU s are to one another, how good the channel conditions are, etc.). Therefore, the rank of the {\em spectrum occupancy matrix} (whose columns each corresponds to the occupancies of the different bands as seen by the corresponding \SU) is likely to be low, and hence, one can exploit such a low-rank property to construct the occupancy matrix from only a smaller number of measurements.
More specifically, each \SU~will only need to sense and report measurements on a small portion of the wideband spectrum (i.e. an \SU~can still use \proposed~to recover band occupancies, but only focusing on one portion of the spectrum). With this, most entries of the spectrum occupancy matrix, will be missing, but these missing entries can be fully recovered by means of low-rank matrix theory~\cite{candes2009exact} (i.e. by formulating an optimization problem whose objective is to minimize the rank of the matrix), as long as the number of observed entries in the spectrum occupancy matrix is at least $O(\alpha^{5/4}r\log \alpha)$ with $r$ as the rank of the spectrum occupancy matrix and $\alpha=\max(\I,\J)$~\cite{candes2009exact}, where $\I$ and $\J$ being the number of narrow-bands and the number of \SU s, respectively.

\subsubsection{{Proposed Low-Rank Matrix Approximation Based Solution}}
When considering high frequencies in mmWave bands, existing low-rank matrix approximation based approaches fail, as the low-rank property is no longer preserved. This is because different \SU s are now very likely to observe completely different occupancy vectors due to high signal path losses. To address this, we propose to leverage two
key facts/observations:
$(i)$ {Though in general, \SU~will not be observing the same band occupancies, close-by \SU s may still observe and report similar occupancy vectors}.
Therefore, if the entries of close-by \SU s are re-arranged in the global spectrum occupancy matrix based on their locations, then the low-rank property will be {\em locally} preserved.
$(ii)$ {User/device density of 5G networks will be high}. The massive nature of 5G devices will be exploited to provide efficient recovery of the occupancy matrix.

\begin{figure}
	\center{
	\includegraphics[width=1\columnwidth]{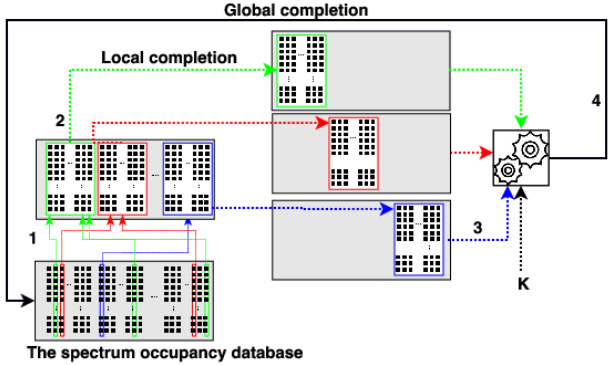}
	\caption{\small {Recovery steps: (1) sub-matrices construction, (2) local low-rank sub-matrices completion, (3) global spectrum matrix recovery.}}
	\label{fig:airmap}}
\end{figure}

The proposed technique consists of the following three steps/components (See Fig.~\ref{fig:airmap}):

\myitemizebegin
{\item {Spectrum sub-matrices construction:}}
The region of interest is divided into $\q$ sub-regions, where the boundary of each sub-region is decided based on how far the highest carrier frequency can be detected using conventional propagation models.
After deciding on the number of sub-regions
and their anchor points (centers),
the \SU s are grouped into clusters with each \SU~being associated with the closest cluster head, whose task is to collect the sensed information from \SU s and reconstruct the sub-matrix, as discussed later.
%
It is worth noting that unlike the approaches proposed for recommendation systems, which use local low-rank matrix approximation such as LLORMA~\cite{lee2016llorma}, the anchor points are constructed independently from the sensing nodes. These depend only on the highest frequency in the mmWave bands and on the size of the region of interest.

{\item {Local low-rank sub-matrices completion:}}
Consider the spectrum occupancy matrix corresponding to a given given sub-region. To reconstruct this sub-matrix, we formulate the problem as an optimization whose objective is to find an unkown matrix  with minimum rank subject to the error with respect the known entries of the matrix.
A key challenge lies in developing an efficient technique for solving this optimization formulation. In this work, we do so by resorting to minimizing the nuclear norm of the matrix as opposed to its rank, which is known to be a convex relaxation of the matrix's rank (i.e., the number of non-zero eigenvalues)~\cite{recht2010guaranteed}.

{\item {Global spectrum matrix recovery:}}
In this step, all constructed sub-matrices are combined to recover the global spectrum occupancy matrix. One challenge this operation faces is that \SU s located at the region edges may have inconsistent measurements (i.e. occupancy information could be different depending on which cluster the \SU~ends belonging to), and hence, without accounting for neighboring decisions, occupancy information may be biased and inaccurate. To help mitigate these boundary issues, we refine the elements of the global spectrum matrix by combining the decisions for a given user using some kernel (similarity) function. The similarity function is a distance-based (the distance of each user from the central of each sub-region. The higher the distance is, the smaller will be the similatrity function.

\myitemizeend

\subsubsection{Further Research Challenges}
One point requires further investigation is to consider applying different similarity functions, and study their impact on the accuracy of occupancy matrix information. Note that the potential of this approach relies on the fact that the number of devices located within each sub-region (e.g. small cell) is high enough, thanks to the high user density feature that 5G is projected to have.

\section{Conclusion}
The shift to mmWave band access has certainly brought great potentials for overcoming spectrum shortage problems, being undoubtedly perceived as one of the key challenges that next-generation networks are facing. However, in order to enable spectrum-efficient communication in these high frequency bands, some key spectrum management challenges need to be overcome.
In this paper, we focus on challenges related to dynamic spectrum access, a key communication paradigm that has been recognized by the various communication technology enablers as key component to next-generation wireless systems. We specifically highlight challenges that arise from the limited hardware capability of transceiver designs, the time- and frequency-varying occupancy/usage of wideband spectrum, and the increased signal path loss due to transmissions occurring at high frequencies. We also propose two techniques that aim to maintain high spectrum efficiency while addressing the time/frequency-variability of spectrum occupancy and the increased path loss challenges, and highlight some research challenges that require future attention to fully unleash the potential of dynamic spectrum sharing at mmWave bands.

\end{document}